\documentclass[twocolumn]{pasj00}
\draft

\begin{document}
\SetRunningHead{K. Wada}{Structure of ISM in ULIRGs}
\Received{}
\Accepted{}

\title{Origin of Warm High-Velocity Dense Gas in ULIRGs}

\author{Keiichi \textsc{Wada}}%
\affil{National Astronomical Observatory of Japan}
\email{wada.keiichi@nao.ac.jp}


%


\maketitle

\begin{abstract}
Possible origins of the molecular absorption discovered in some ULIRGs are
investigated, based on a 3-D hydrodynamic model of
star-forming interstellar gas in a galactic central region.
The blue-shifted, warm ($\sim 200-300$ K), dense ($>10^6$ cm$^{-3}$) 
molecular gas suggested by CO absorption in IRAS 08752+3915 could be 
caused by the innermost region of the inhomogeneous inter-stellar medium (ISM)
around a supermassive black hole. 
The infrequent observations of the dense gas with absorption in ULIRGs and
Seyfert 2 galaxies could simply suggest that the high-density regions
occupy only a very small volume fraction of the obscuring material.
This is naturally expected if the inhomogeneous structure of the ISM is caused by
non-linear development of instabilities. 
The model predicts a turbulent velocity field in
the obscuring material, therefore 
blue- and red-shifted gases should be observable with nearly the same probability
for the large enough statistical samples.
 \end{abstract}

\section{Introduction}

Active Galactic Nuclei (AGNs) are believed to be highly obscured by
dusty dense gas, and this contributes to huge radiative energy in some
ultra-luminous infrared galaxies (ULIRG) \citep{imani00}. The central
engines of type-2 Seyfert galaxies are also likely obscured by optically
thick molecular gas, for which a torus-like geometry is often assumed.
However, detailed structures of the ISM in deep centers of ULIRGs and
Seyfert galaxies are still unclear.
There are some indirect ways to infer their geometry and distribution
using X-ray spectroscopy (e.g., \cite{lev01}), optical/infrared spectral
energy distribution (SED) in comparison with model SED assuming the
geometry of the torus (e.g., \cite{pier93,elit06, fritz06}), and
optical/infrared absorption lines (e.g., \cite{spoon04,imani06, imani07,
lev07}).  Lutz et al. (2004) searched the 4.7 $\mu$m fundamental
ro-vibrational band of CO in 31 bright local AGNs, but found no clear
signature of absorption features, even in a Compton-thick Seyfert 2
nucleus, like NGC 1068.  Recently, the {\it Spitzer} Infrared Spectrograph
(IRS) revealed vibration-rotation absorption bands of gaseous
C$_2$H$_2$, HCN, and CO$_2$ as well as silicate absorption toward deeply
obscured (U)LIRG nuclei \citep{spoon04, lahuis06, lev07}.
The absorption lines of C$_2$H$_2$ and HCN suggest the presence of warm
($T_g \simeq 200-700$ K) and dense ($n_{\rm H} > 3\times 10^6$
cm$^{-3}$) gas \citep{lahuis06}.  They suggest that this gas occupies
only a small fraction of the nuclear region ($\sim$ 0.01 pc) near the
intrinsic mid-infrared source. The CO absorption features observed in
ULIRGs are not resolved into individual lines, except IRAS 08572+3915,
therefore the kinematics and structure of the absorbed material are
still an open question.


\citet{gebal06} have found a broad CO absorption line toward IRAS
08572+3915 NW\footnote{Optical classification of this galaxy is LINER \citep{veil99}. 
\cite{imani00}, based on infrared spectroscopy, claimed that the AGNs
of this galaxy and many other non-Seyfert ULIRGs are deeply buried in dusty gas.} using UKIRT/CGS4.  The observations revealed the following
features: 1) There are several blue-shifted components with $-160 \pm
25$ km s$^{-1}$ for the high $J$ lines ($P(6), P(8), ...$), and $-150
\pm 25$ km s$^{-1}$ and $-50 \pm 25$ km s$^{-1}$ and for the low $J$
lines ($P(1), R(1)$ and $R(2)$).  2) The hydrogen column density
is approximately $1.5\times 10^{22}$ cm$^{-2}$.  3)
Mean temperature of the warm blue-shifted component is about 200 K.

\citet{shira07} have also detected the CO absorption in IRAS 08572+3915
NW using Subaru/IRCS.  Their results
are consistent with those of \citet{gebal06}, showing  a blue-shifted
component (-160 km s$^{-1}$) with the temperature $273$ K and cold gas
at $27$ K at the systematic velocity, assuming LTE.  They also claim
that there is a red-shifted component (+100 km s$^{-1}$) seen in high $J$
($J > 4$) lines, implying higher temperature gas ($\sim 700$ K).
Column density of the warm component is $3\times 10^{22}$
cm$^{-2}$.

These observations suggest that the absorption features do not simply
arise from a smooth rotating molecular torus, which has been often
postulated to explain the type-1 and type-2 AGNs. However, it is hard to
determine the geometry and internal structures of the ISM in the central
region of the galaxy from this observational
information alone. Theoretical models used for SED fitting are
phenomenological without kinematical information, therefore we cannot
compare them with the absorption line observations.


In this paper, we investigate how the observational features suggested
by the CO absorption in IRAS-08572+3915 NW can be understood in the
context of a three-dimensional hydrodynamic model of the ISM around a
supermassive black hole (SMBH) with nuclear starbursts \citep{wad02}
(hereafter WN02).  This is currently a unique model of the obscuring
material around a SMBH on a several tens pc scale, characterized by a
highly inhomogeneous, multi-phase, and turbulent ISM with a globally
stable, geometrically thick structure.  Three-dimensional radiative
transfer calculations for this model revealed that CO luminosity
distribution is also highly non-uniform \citep{wad05}.  Although this is
not a confirmed theoretical model applicable to the obscuring matter in
all types of AGNs, it is worth verifying whether the features observed
in the CO absorption lines in the ULIRG can be explained by the full 3-D
hydrodynamic models.


\section{Model and Analysis}
\subsection{A Hydrodynamic Model of  the ISM around a SMBH}
In the WN02 model, mass, momentum, and energy conservation equations
with the Poisson equation are numerically solved with energy feedback from
supernovae in a fixed gravitational potential.
A rotating gas disk in a time-independent
spherical potential
is solved by 3-D hydrodynamic code. 
The mass of the BH is $M_{\rm BH} = 10^8 M_\odot$.
We also assume a cooling function $\Lambda(T_g) $ $(5 K < T_g < 10^8
{\rm K})$ with solar metallicity, heating due to
photoelectric heating, and energy feedback
from SNe.  We assume a uniform UV radiation field 
ten times larger than the local UV field.

The hydrodynamic part of the basic equations is solved by AUSM
 (Advection Upstream Splitting Method) \citep{liou03}.
(See details in \citet{wada01, wad01b}).
We use $256^2 \times 128$ 
Cartesian grid points covering a $64^2\times 32$ pc$^3$ 
around the galactic center (i.e. the
spatial resolution is 0.25 pc).  The Poisson equation is solved 
using the fast Fourier transform and the convolution method.
The initial condition is an axisymmetric and rotationally
supported thin disk (the scale height 
is 2.5 pc) with a uniform radial density profile and a total gas mass of $M_g = 5\times 10^7 M_\odot$.
Random density and temperature fluctuations, which are 
less than 1 \% of the unperturbed values, 
are added to the initial disk.

Supernova (SN) explosions are assumed to occur at random positions
on the disk plane.
The average SN
rate is $\simeq$ 0.8 yr$^{-1}$.
The energy of $10^{51}$ ergs is instantaneously
injected into a single cell as thermal energy. 
Thus the three-dimensional evolution of 
blast waves driven by the SNe in an inhomogeneous and non-stationary
medium with a global rotation is followed explicitly, taking into account the radiative cooling.

\begin{figure}
   \begin{center}
     \FigureFile(80mm,80mm){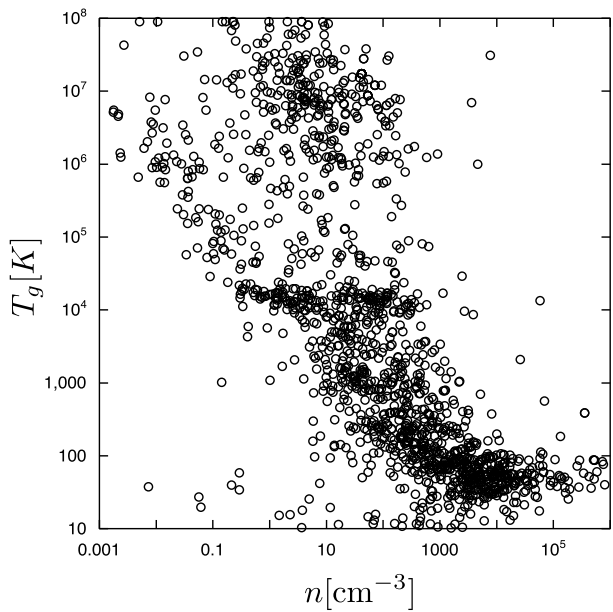}
   \end{center}
   \caption{Density-Temperature diagram of the ISM around a supermassive
black hole with energy feedback from supernovae. The open circles 
represent 500 sampled point in the torus. }
\label{fig:phase}
 \end{figure}

WN02 showed that a globally stable concave ``torus'', in which the gas is
highly turbulent and inhomogeneous, is formed (See also figure 6).
Figure \ref{fig:phase} is a phase-diagram of the ISM in the torus
 at a quasi-steady state.
Three dominant phases present: Hot gases around $T_g \sim 10^7$ K 
caused by SNe, warm gases around $T_g \sim 10^4$ K, and 
cold, dense gases at $T_g < 100-1000 $ K. 
In terms of the observed CO absorption in IRAS 08572+3915, 
we are interested in kinematics and spatial location of the cold, dense media.

The size of the present torus model ($r \simeq 30$ pc) is 
much larger than the size ($\sim 2$ pc) suggested by near- and mid-infrared 
high-resolution observations of type-2 Seyfert, e.g. NGC 1068
\citep{jaffe04,wittk04}. 
This difference actually reflects a fact that 
near/mid-infrared flux is originated from 
inner a few pc region of the dusty material around the AGN.
In fact, a recent 3-D radiative transfer model assuming 
clumpy dust tori for NGC 1068 \citep{hoenig06} suggests that 
the outer radius of the torus that fits SED of NGC 1068
is 56 pc, while their H-, K-, and N-band images show their radii
are 2.0-2.7 pc.  In terms of ULIRGs, 
\citet{lev07} suggested observed deep absorption features 
in mid-infrared indicate
obscuration on scales of a few 100 pc. 
Probably the obscuring material in the central region of
AGNs and ULIRGs is extended from a sub-pc to a sub-kpc scales
with various physical/chemical conditions and structures.

\subsection{Analysis of the Hydrodynamic Results}

In order to find the blue- and red-shifted warm, dense gases suggested
by the CO absorption line in the hydrodynamic model, we sample points with the following conditions
from the  2000 randomly selected points in a half (i.e. $z\geq 0$) the
``torus''.
\begin{enumerate}
 \item Number density is $n \geq 10^6$ cm$^{-3}$
 \item Column density from the nuclear source to the observer $N \geq 10^{23}$ cm$^{-2}$
 \item Temperature is  $10 \leq T_g \leq 1000$ K
\end{enumerate}

 \begin{figure}[h]
   \begin{center}
     \FigureFile(80mm,80mm){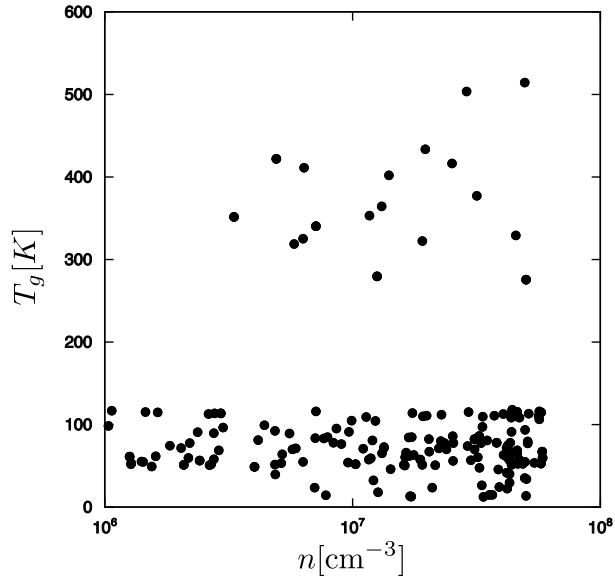}
   \end{center}
   \caption{Density and temperature of gases which are candidates
for the absorbed gas.}\label{fig:density-temp}
 \end{figure}

\begin{figure}[h]
   \begin{center}
     \FigureFile(80mm,80mm){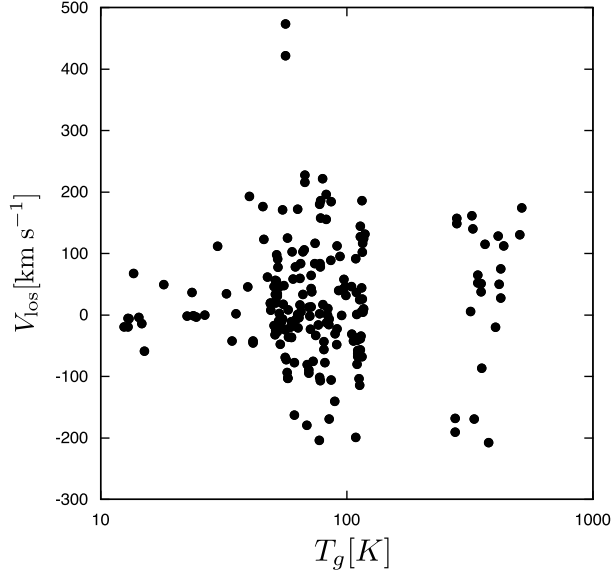}
   \end{center}
   \caption{Same as figure \ref{fig:density-temp}, but temperature vs. line-of-sight velocity.}\label{fig:vlos-temp}
 \end{figure}

\begin{figure}[h]
   \begin{center}
     \FigureFile(80mm,80mm){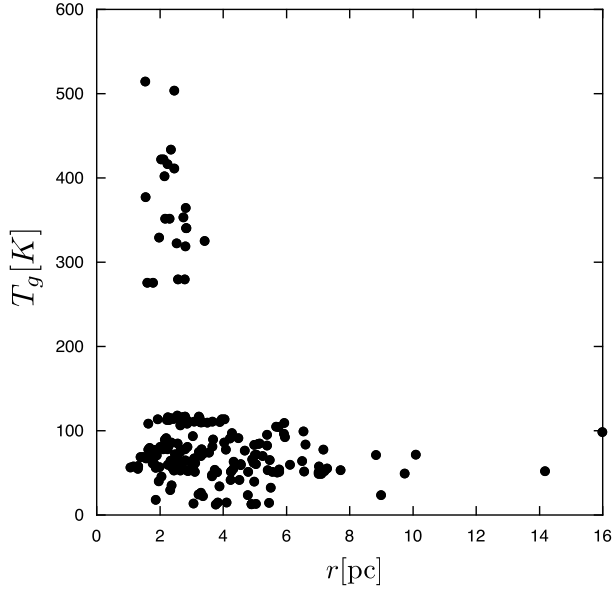}
   \end{center}
   \caption{Radial distribution of the selected points.}\label{fig:r-temp}
 \end{figure}

\begin{figure}[h]
   \begin{center}
     \FigureFile(80mm,80mm){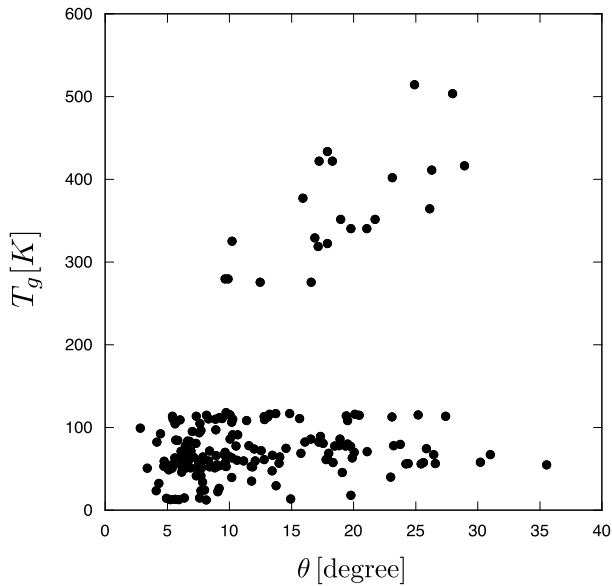}
   \end{center}
   \caption{Same as figure \ref{fig:r-temp}, but for temperature
  vs. position
angle $\theta$. $\theta=0$ is on the equatorial plane of the torus.}\label{fig:theta-temp}
 \end{figure}

Figure \ref{fig:density-temp} shows number density and temperature of
the selected points. It is clear that two components exist:  cold gas with $  T_g  \lesssim
100$ K and warm gas with $ 300 \lesssim T_g \lesssim 500$ K.
 Figure \ref{fig:vlos-temp} shows the line-of-sight velocity of the
 selected points, $V_{\rm los}$, as a function of their temperature.
 Most of the cold and warm components are distributed in the range of
 $|V_{\rm los}| \lesssim 200 $ km s$^{-1}$.
Spatial distribution of the selected points are shown in figures \ref{fig:r-temp} and \ref{fig:theta-temp}.
The warm components are distributed around $2 \lesssim r \lesssim 4$ pc 
 from the galactic center, and they are located just above the equatorial plane, i.e. 
$10 \lesssim |\theta| \lesssim 30^\circ$. Temperature of the warm
component
is higher for larger $|\theta|$, i.e. closer to the surface of the ``torus''.
The cold dense gases are located $2 \lesssim r \lesssim 10$ pc and
near the disk plane ($|\theta| \lesssim 10^\circ$).

 \begin{figure}[h]
   \begin{center}
     \FigureFile(80mm,80mm){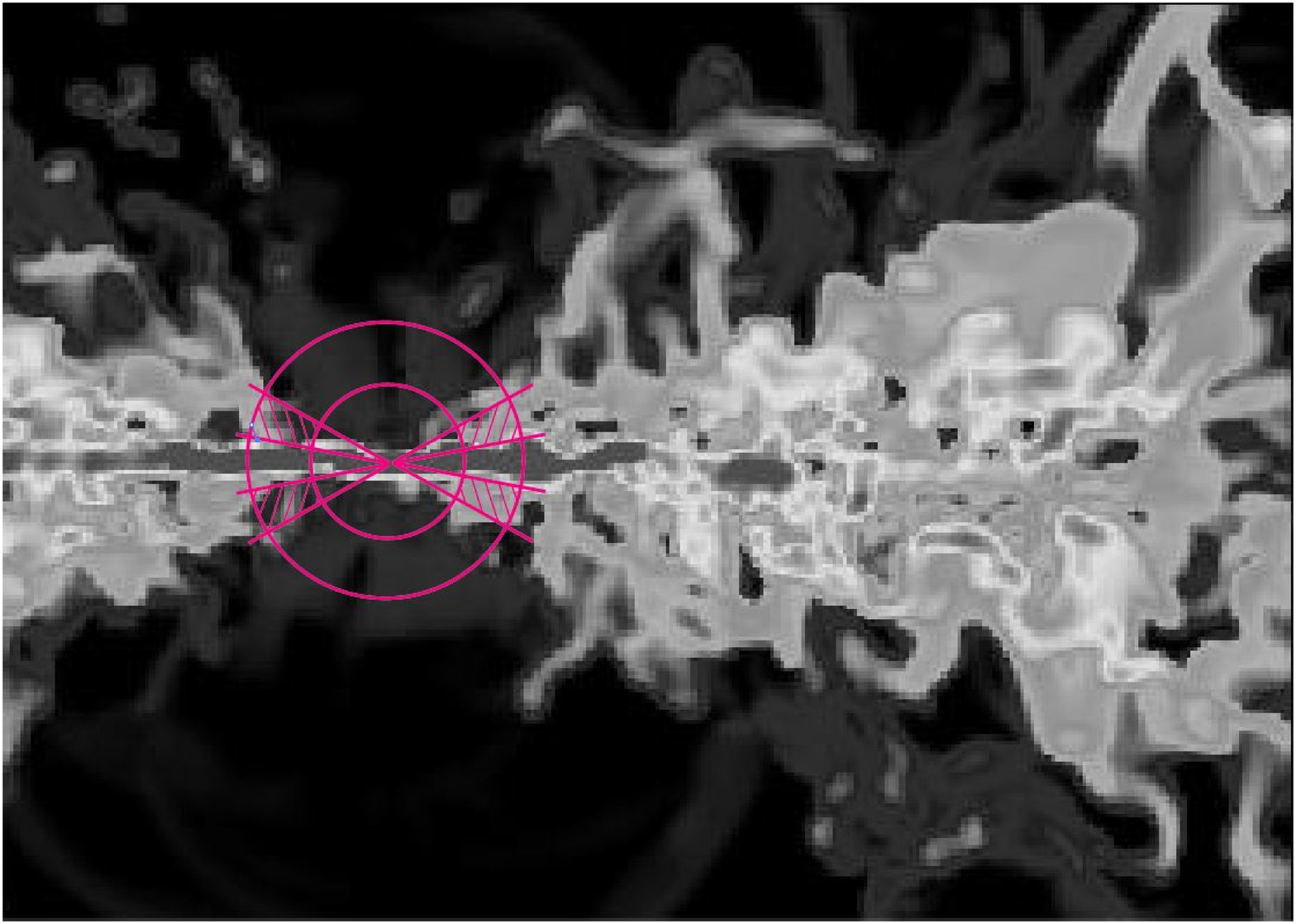}
     \FigureFile(100mm,100mm){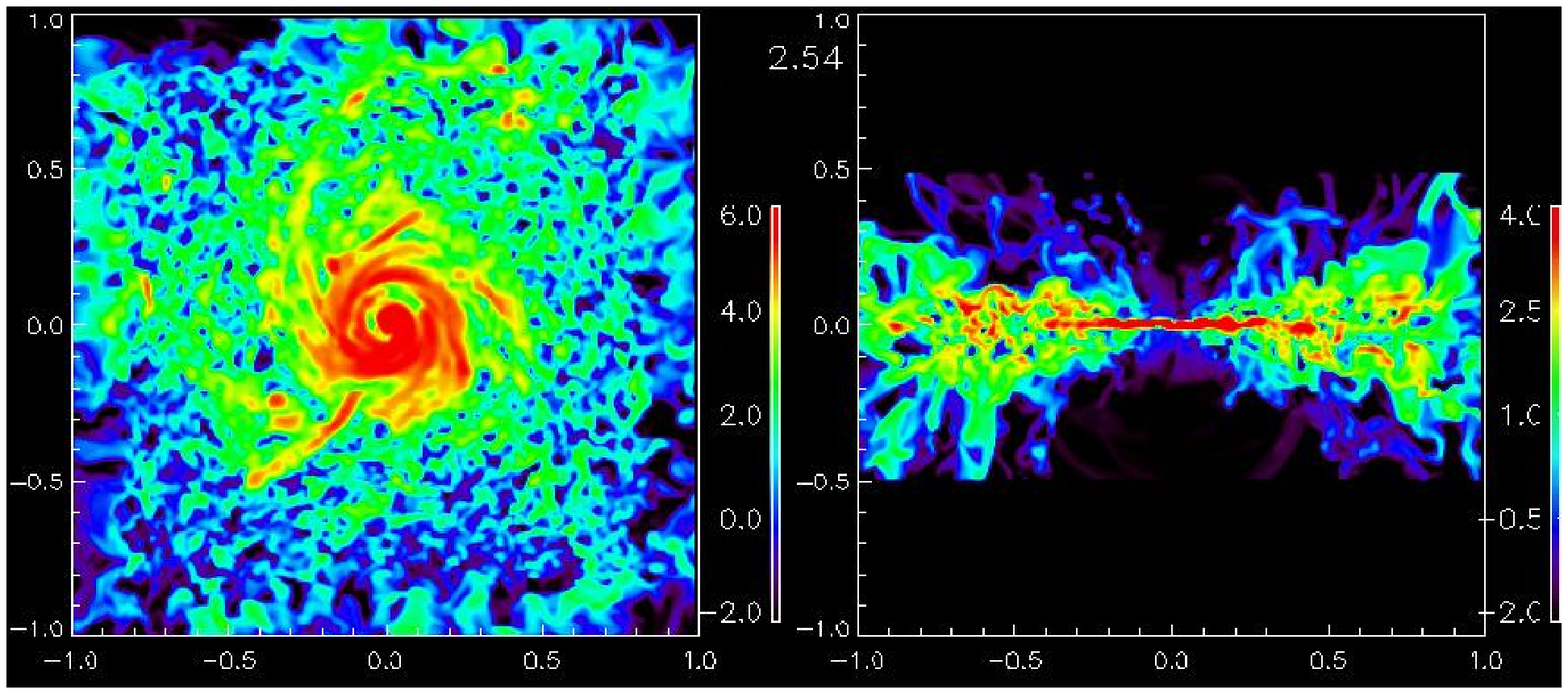}
   \end{center}
   \caption{(upper panel) Possible location (the hatched region) of the warm gas
  clumps suggested by CO absorption observations.
See the lower panel, in which log-scaled gas density is shown 
at x-y and x-z planes (unit of density and scale are $M_\odot {\rm pc}^{-3}$ 
and 32 pc).}
\label{fig:torus}
 \end{figure}

In summary, the absorbed material with $N > 10^{23}$ cm$^{-2}$, $n >
10^6$ cm$^{-3}$, $|V_{\rm los}| = 100-200$ km s$^{-1}$, and $T_g \simeq
300-500$ K sits at the innermost region of the inhomogeneous torus as
shown in figure \ref{fig:torus} (the hatched regions).  For the gas near
the equatorial plane, the column density is too high to be observed as
an absorption line for the continuum IR source.  Therefore, if the ISM
in the central region of ULIRGs, like IRAS 08572+3915, is similar to the
structure shown in figure \ref{fig:torus}, the viewing angle to observe
the warm and dense gases as absorption is restricted.  If this is the
case, we can expect to observe high velocity, warm components as well as
low velocity, cold components as absorption against the background
continuum source, whose size is smaller than $1$ pc.  The high density
gases in those regions have various velocity components, as shown in
figure \ref{fig:vlos-temp}.  Thus it is not necessary to observe only
blue-shifted gas.  In fact, \citet{shira07} also found a red-shifted gas
component (+100 km s$^{-1}$) in high-$J$ transition, suggesting 
gas warmer than the blue-shifted component.

One should note that geometry of the ``torus'' in the present model depends on
the supernovae rate, total gas mass, and the mass of a SMBH, 
since it is determined from a balance between energy input and
dissipation rates under the effect of gravitational
potential (WN02).
Therefore, the absolute locations (i.e. $r$ and $\theta$) of the warm, dense gas
around an AGN should depend on these parameters.

\section{Discussion}

\subsection{Volume-Filling Factor of the Warm, Dense Gas}

In the present model of the torus, the high density ($> 10^6$ cm$^{-3}$)
gases have both red- and blue-shifted velocity for a line-of-sight.
Since a volume fraction of these dense gases 
is expected to be very small (see discussion below), and
the background continuum is emitted from a
small region in the vicinity of the central engine, 
the chance to observe a dense blob with a particular velocity component
by an absorption line should be very small.
For example, as shown in figure \ref{fig:vlos-temp},  the blue-shifted component
around $-200$ km s$^{-1}$ exists in about 1/500 of the total volume of the
torus whose radius is about 30 pc.
Therefore, one possible explanation about the observed blue-shifted component
is that the line-of-sight toward the AGN
happens to graze the high-density clump which is approaching us.
If this is the case, since the motion in the torus is highly turbulent (WN02), 
we have nearly equal chance of observing red-shifted warm gas, too. 
The red-shifted warm gas found by \citet{shira07} may be
explained by this picture, but observed samples are still too
small to further compare with the model.

Recently \citet{wad07} proposed a simple statistical theory of
the structure of the inhomogeneous gas disk in galactic disks, which is caused by 
non-linear development of gravitational and thermal instabilities.
Using three-dimensional hydrodynamic simulations, they show
that the probability distribution function (PDF) of density $P(\rho)$
in a globally stable, inhomogeneous ISM 
is well represented by a single log-normal function over a wide density
range, that is 
\begin{eqnarray}
P(\rho)d\rho = \frac{1}{\sqrt{2\pi}\sigma}
\exp{\left[
-\frac{
\ln (\rho/\rho_0)^2
}{
2\sigma^2
}
\right] } d \ln \rho,
\end{eqnarray}
where $\rho_0$ is the characteristic density and $\sigma$ is the
dispersion.
The simulations show that the dispersion of the log-normal PDF is larger for
more massive systems. 
Using the PDF, it is straightforward to calculate
a volume fraction of high density gas ($f_c$) above a given critical density ($\rho_c$):
\begin{eqnarray}
f_c (\rho_c, \sigma) = \frac{1}{2}\left(1 - {\rm Erf}\left[
\frac{\ln (\rho_c/\rho_0) - \sigma^2}{\sqrt{2} \sigma}
\right]\right)\;.
\label{eq: fc}
\end{eqnarray}
The dispersion $\sigma$ is then related to the average density,
$\bar{\rho}$, i.e.
\begin{eqnarray}
\sigma^2 = 2 \ln \left(\frac{\bar{\rho}}{\rho_0}   \right).
\label{eq: sigma}
\end{eqnarray}
It is also suggested that characteristic density $\rho_0$ is not very sensitive for changing
the total gas mass.
Suppose the column density toward the continuum source is
$10^{22}$ cm$^{-2}$ and the size of the `torus' is 30 pc, 
the average density would be $\bar{\rho} \simeq 10^2$ cm$^{-3}$.
Therefore, using equation (\ref{eq: sigma}), 
we have $\sigma \simeq 3.0$ for $\rho_0 = 1$ cm$^{-3}$.
This leads to $f_c = 9.5\times 10^{-4}$ and $f_c = 5.3\times 10^{-5}$ for
$\rho_c = 10^6$ cm$^{-3}$ and $10^7$ cm$^{-3}$, respectively.
With this small volume-filling factor of the high-density gas in an obscuring
material, it might be reasonable to expect
there is a small chance of detecting
a high-density clump with a
particular velocity by absorption lines.

\subsection{Difference between ULIRG and Seyfert1,2}

\citet{hao07} observed 196 AGNs and ULIRGs using {\it Spitzer}/IRS, and
found that quasars are characterized by silicate features in emission
and Seyfert2s are dominated by weak silicate absorption,
while ULIRGs are characterized by strong silicate absorption.
\citet{lev07} suggest that this difference in
near-infrared spectra reflects a
different geometry of the 
dusty material. They claim that  clouds dominate
the obscuration of Seyfert 1s and 2s, while on the other hand,
central sources of ULIRGs are obscured mainly by smooth,
geometrically and optically thick material with 
a steep temperature gradient.
According to \citet{lahuis06},  $N_{\rm H_2}$ for the nucleus of
IRAS 08572$+$3915 estimated by infrared spectra obtained by {\it
Spitzer}  is $N_{\rm H_2} \simeq 1.5\times 10^{23}$ cm$^{-2}$.
If the scale of the obscuring material is about 30 (100) pc, the
average number density is $n \simeq 1.5\; (0.5)\times 10^{3}$ cm$^{-3}$.
Using the log-normal density PDF in \S 3.1, we can estimate that
the volume filling factor of those gases is about 20 (30) \%,
suggesting that the dusty gas that contributes to the silicate
absorption is more smoothly distributed in
the obscuring material than the warm, dense gas found by CO absorption.

Our result presented here is that at least the CO absorbed feature of
IRAS 08572+3915 is not inconsistent with a model characterizing a dense
clumpy medium around an AGN.  Since the velocity field of the ISM is
highly turbulent in the model, if individual high density clumps are not
resolved, a broad absorption feature originating in dense gases with
various velocity components can be expected in the obscuring material.
If this is the case, the reason why CO absorption is not observed in
Seyfert-2 \citep{lutz04} can be understood by the low filling factor of
the warm, dense gas in the obscuring material.  As discussed in \S 3.1,
if average gas density of the obscuring material around AGNs is smaller
in Seyfert galaxies than in ULIRGs, a fraction of high-density clumps
becomes small.  This seems to be inconsistent with the result by
\citet{lev07}, but probably dusty ``clouds'' in the obscuring material
responsible for the near/mid-infrared SED are much more diffuse than
those causing the CO absorption (see discussion above). The ISM around AGNs in ULIRGs and
Seyfert galaxies are more or less inhomogeneous, and if so its density
PDF is expected to be approximately log-normal.  Therefore, it is not
surprising, if the ``smooth'' distribution of dusty gas that is
preferable to the SED fitting would include higher-density clumps with a
smaller volume-filling factor.


\subsection{Limitations of the Current Model and Other Possibilities}



In the present hydrodynamic model, heating due to the radiation 
from the AGN to dust is not taken into account. 
Temperatures of optically thick clouds at a distance of a few pc from
an AGN with $L=10^{12}L_\odot $would be 300-1200 K \citep{elit06}.
Therefore the warm clouds near the surface of the 
clumpy ``torus'' might be warmer than $300-500$ K, 
which is a result of the present model.
Effects of X-rays are not included in the model either, which 
would be especially important for chemistry of molecules
\citep{malo96,meij07}. For highly inhomogeneous media like
those seen in figure \ref{fig:torus}, hard X-rays could affect not 
only chemistry of the gas in the vicinity of the central source,
but also the clouds inside the ``torus'', because 
the column density and the ionization rate are not simply related to the distance
from the central source (WN02).
The effects of X-rays on the abundance and line intensities 
of CO and other molecules are investigated using the present 
hydrodynamic model and 3-D non-LTE radiative transfer calculations
(Yamada, Wada, \& Tomisaka, in preparation).

As an alternative interpretation of 
the blue-shifted warm gas could be originated in outflows from
an accretion disk around a supermassive black hole.
Using two-dimensional radiation-hydrodynamic simulations, 
\citet{ohsuga07} found that  an outflow is driven by radiation force
due to a luminous accretion disk. However, the outflow velocity is
extremely large ($\sim 0.1 c$), and its temperature is $\sim 10^7$ K,
therefore the outflow from the accretion disk 
itself does not explain the warm, blue-shifted
gases discussed here. It might be possible however that, 
the hot outflow gas propagates outward, and it then
cools and forms warm, dense cloudlets by the thermal instability
at a few pc from the accretion disk. Unfortunately long-term evolution
of the outflow from the accretion disk and its interaction with the ISM
on several tens pc scale is still an open question.

\section{Conclusion}

Based on the analysis presented here, we could conclude that the
blue-shifted, warm, dense molecular gas in IRAS 08752+3915 NW suggested
by recent high-resolution observations by Subaru/IRCS can be caused by
high-density clumps in the inhomogeous inter stellar matter around a
supermassive black hole. The small possibility of observing the CO
absorption in ULIRGs and Seyfert 2 galaxies could be simply interpreted
as indicating that the high density regions occupy a very small volume
fraction ($< 10^{-3}-10^{-4}$) in the obscuring material with a
turbulent velocity field.  This is naturally expected if the
inhomogeneous material is caused by non-linear development of
instabilities and random processes (e.g. stochastic explosions of
supernovae and interaction with inhomogeneous ISM) in
a globally quasi-stable system.

However, this model is still a presumption until the absorbed features
in near infrared spectra of many ULIRGs and Seyfert 2 galaxies are
resolved and details of the kinematics of the obscuring material are
clarified by future observations.  It is also necessary to improve
numerical models of the ISM in the central region of active galaxies by
taking into account more realistic treatment of chemical reactions of
molecular gas and radiative transfer in terms of AGNs and star forming
regions.

%
\vspace{0.5cm}
%
The author is grateful to Takao Nakagawa, Mai Shirahata, and Masa Imanishi
for fruitful discussions. The anonymous referee's comments were also helpful.
Numerical computations were carried out on a Fujitsu VPP5000 at NAOJ.
KW is supported by Grant-in-Aids for Scientific Research
[no. 16204012] of JSPS.



\begin{thebibliography}{}
\bibitem[Elitzur(2006)]{elit06} Elitzur, M.\ 2006, New Astronomy Review, 50, 728
\bibitem[Fritz, Franceschini, \& Hatziminaoglou (2006)]{fritz06} Fritz, J., Franceschini, A., \& Hatziminaoglou, E. 2006, MNRAS, 366, 767
\bibitem[Geballe et al.(2006)]{gebal06} Geballe, T.~R., Goto, M., Usuda, T., Oka, T., \& McCall, B.~J.\ 2006, \apj, 644, 907 
\bibitem[Hao et al.(2007)]{hao07} Hao, L., Weedman, D.~W., 
Spoon, H.~W.~W., Marshall, J.~A., Levenson, N.~A., Elitzur, M., 
\& Houck, J.~R.\ 2007, \apjl, 655, L77 
\bibitem[H{\"o}nig et al.(2006)]{hoenig06} H{\"o}nig, S.~F., 
Beckert, T., Ohnaka, K., \& Weigelt, G.\ 2006, \aap, 452, 459 
\bibitem[Imanishi \& Dudley(2000)]{imani00} Imanishi, M., \& 
Dudley, C.~C.\ 2000, \apj, 545, 701 
\bibitem[Imanishi et al.(2006)]{imani06} Imanishi, M., Dudley, 
C.~C., \& Maloney, P.~R.\ 2006, \apj, 637, 114 
\bibitem[Imanishi et al.(2007)]{imani07} Imanishi, M., et al. ApJ in press (astro-ph/0702136)
		       \aap, 404, 495 
\bibitem[Jaffe et al.(2004)]{jaffe04} Jaffe, W., et al.\ 2004, 
\nat, 429, 47 
\bibitem[Lahuis et al.(2007)]{lahuis06} Lahuis, F., et al.\ 
2007, \apj, 659, 296 
\bibitem[Levenson, Weaver, \& Heckman(2001)]{lev01} Levenson, N.~A., Weaver, K.~A., \& Heckman, T.~M.\ 2001, \apj, 550, 230 
\bibitem[Levenson et al.(2007)]{lev07} Levenson, N.~A., Sirocky, M.~M., Hao, L., Spoon, H.~W.~W., Marshall, J.~A., Elitzur, M., \& 
Houck, J.~R.\ 2007, \apjl, 654, L45 
\bibitem[Liou \& Steffen(1993)]{liou03} Liou, M., Steffen, C., 1993,				    J.Comp.Phys., 107,23
\bibitem[Lutz et al.(2004)]{lutz04} Lutz, D. et al. \aap, 426, L5
\bibitem[Maloney et al.(1996)]{malo96} Maloney, P.~R., 
Hollenbach, D.~J., \& Tielens, A.~G.~G.~M.\ 1996, \apj, 466, 561 
\bibitem[Meijerink et al.(2007)]{meij07} Meijerink, R., 
Spaans, M., \& Israel, F.~P.\ 2007, \aap, 461, 793
\bibitem[Ohsuga(2006)]{ohsuga07} Ohsuga, K.\ 2006, \apj, 659, 205
\bibitem[Pier \& Krolik(1993)]{pier93} Pier, E.~A., \& Krolik, 
J.~H.\ 1993, \apj, 418, 673 
\bibitem[Shirahata et al.(2007)]{shira07} Shirahata, M., Nakagawa, T.,
				    Goto, M., Usuda, T., Suto, H., \&
				    Geballe, T.R., 2007, submitted to ApJ
\bibitem[Spoon et al.(2004)]{spoon04} Spoon, H.~W.~W., et al.\ 
2004, \apjs, 154, 184 
\bibitem[Veilleux et al.(1999)]{veil99} Veilleux, S., Sanders, D.~B., \& Kim, D.-C.\ 1999, \apj, 522, 139 
\bibitem[Wada(2001)]{wada01} Wada, K. 2001, \apjl, 559, L41
\bibitem[Wada \& Norman(2001)]{wad01b} Wada, K.\ \& Norman, C.\ A.\ 2001, \apj, 547, 172
\bibitem[Wada \& Norman(2002)]{wad02} Wada, K.~\& Norman, C.~A.\ 2002,				    \apjl, 566, L21 (WN02)
\bibitem[Wada \& Norman(2007)]{wad07} Wada, K.~\& Norman, C.~A.\ 2007,
		       \apj, 660 no.2, in press (astro-ph/0701595)
\bibitem[Wada \& Tomisaka(2005)]{wad05} Wada, K.~\& Tomisaka, K. 2005,
		       \apj, 619, 93
		       Meurer, G., \& Norman, C.~A.\ 2002, \apj, 577,
		       197 
\bibitem[Wittkowski et al.(2004)]{wittk04} Wittkowski, M., 
Kervella, P., Arsenault, R., Paresce, F., Beckert, T., \& Weigelt, G.\ 
2004, \aap, 418, L39 
\end{thebibliography}
\end{document}